\documentclass[letterpaper,11pt]{revtex4}
\usepackage{graphicx}
\usepackage{bm}

\def\beq{\begin{eqnarray}}
\def\eeq{\end{eqnarray}}
\def\be{\begin{equation}}
\def\ee{\end{equation}}

\begin{document} 

\title{\bf On the behaviour of random K-SAT on trees}

\author{Supriya Krishnamurthy$^{(1,2)}$ and Sumedha $^{(3)}$}

\affiliation{(1): Department of Physics, Stockholm University, SE- 106 91, Stockholm, Sweden \\ (2): School of Computer Science and Communication, KTH, SE- 100 44 Stockholm, Sweden \\ (3): National Institute of Science Education and Research, Institute of Physics Campus, Bhubaneswar, Orissa- 751 005, India}

\date{\today}
\begin{abstract}
We consider the K-satisfiability problem on a regular $d$-ary rooted tree.
For this model, we demonstrate how we can calculate in closed form, 
the moments of the total number of solutions as a function of $d$ and $K$, 
where the average is over all realizations, for a fixed assignment 
of the surface variables. We find that different moments 
pick out different 'critical' values of $d$, below which they diverge 
as the total number of variables on the tree $\rightarrow \infty$ 
and above which they decay.  We show that K-SAT on 
the random graph also behaves similarly.
We also calculate exactly the fraction of instances that have solutions
for all $K$. On the tree, this quantity decays to $0$ 
(as the number of variables increases) for any $d>1$. However 
the recursion relations for this quantity have a non-trivial 
fixed-point solution which indicates the existence of a different
transition in the interior of an infinite rooted tree.
\end{abstract}

\maketitle

\section{Introduction}

Constraint satisfaction problems (CSP) are problems in which a set of
variables, which can take values in a specified domain, 
have to satisfy a  number of constraints. Each constraint usually 
restricts the values that a subset of the variables  can take.
The problem then is to find an assignment of the variables that satisfies
all the constraints. The K-satisfiablity problem (K-SAT) is an important 
example of a CSP. In this problem, the variables are 
considered Boolean, taking values True or False, and each constraint
is in the form of a clause, which restricts the values of $K$ 
variables at a time, disallowing $1$ out of the $2^K$ 
possible values that these $K$ variables can take together.

Satisfiablity has been a fundamental problem for almost forty years 
in computer  science. It is known that as soon as there are 
clauses which restrict the values of $K \geq 3$ variables, this 
problem is NP -complete \cite{cook}, {\it i.e.}, 
potential solutions can be verified easily for correctness, but
finding a solution can take exponential time in the worst case.
%there exist 
%instances for which a solution cannot be found by any known algorithm
%which is polynomial-time in $N$ (the number of variables 
%involved in the problem). 
In addition, being NP-complete, 
should a polynomial-time algorithm be found for solving SAT,
it is also possible to adapt it to solve in polynomial-time 
all problems in NP.

The version of the K-SAT that we are interested in, in this paper
is the random K-SAT, which has been very well investigated 
in the past few years. In the random K-SAT one looks at the 
ensemble of randomly generated
logical expressions, where  each logical expression or formula 
is an AND of M
clauses. Each clause consists of an OR of $K$ Boolean literals 
(a literal being one of the $N$ variables or its negation), chosen
randomly from a set of $N$ Boolean variables. As the ratio
$\alpha=M/N$  increases it becomes harder to find satisfiable
assignments for all the $N$ variables that can satisfy  the 
logical expression of $M$ clauses. One of the questions of interest is hence 
if there exist a $\alpha_{c}$, beyond which in the limit of $M
\rightarrow \infty$ and  $N \rightarrow \infty$,  no satisfying
assignments exist.

Numerical experiments have shown in the past that 
if one studies the probability that a randomly chosen formula having 
$M = \alpha N$ clauses is satisfiable, the probability approaches $1$ for 
$\alpha < \alpha_c(K)$ and vanishes for $\alpha > \alpha_c(K)$ when 
$ N \rightarrow \infty $ \cite{msl,kirkpatrick}.
The existence of a sharp transition (the solvability transition)
is intrinsic to the problem and not an artifact of any particular algorithm 
(in \cite{friedgut} the existence of a possibly $N$-dependent transition 
is proved for any $K$), 
but its location has not been determined rigorously with the exception 
of $K=2$ \cite{chvataletal}. 
There are however several rigorous bounds 
on $\alpha_c (K)$,  both upper and lower 
(see \cite{achlioptas_bounds} for a review of these). 
In addition, powerful methods from statistical physics, 
taking advantage of the connection 
of this problem to the theory of mean field spin glasses, 
have been used to conjecture values for the threshold 
that seem to be very accurate 
numerically \cite{mezard-science,mmz}.

The above problems are all originally defined on random graphs, 
where the presence of large loops makes the problem hard to solve exactly.
Hence solving the problem on trees or locally tree-like graphs 
has played an important role in elucidating the nature of the 
solvability transition as well as other phase transitions present in the 
problem. Infact the methods from statistical physics assume the absence 
of correlations between some random variables, which is equivalent to 
solving the K-SAT problem on a locally 
tree-like graph. In addition, it has also been shown that 
certain problems on a tree, in particular the 
tree-reconstruction problem \cite{mezard-montanari}, become equivalent to the 
spin glass problem on a random graph.

In this paper  we study the K-SAT problem on a regular $d$-ary rooted tree
in which every vertex (except the leaves) has exactly $d$ descendents. The values that the
surface nodes (or leaves) take on the tree, are fixed. For a given assignment
of the surface nodes, and a given realization of clauses on the 
tree, one can ask how many assignments of the
variables on the tree are solutions. 
For such a tree graph, we can exactly calculate the moments of 
this quantity averaged over all realizations. The behaviour of the 
moments is similar to that on the random graph, and we find that
there is a different ``critical'' point associated with each moment
as in the random energy model considered by Derrida \cite{derrida}.
We also calculate exactly the fraction of realizations 
that have solutions, for  a tree of any size. This property too 
shows a similarity to the behaviour of the model on a random graph, since there is a  value of $d$ above 
which the fraction of realizations that have solutions $\rightarrow 0 $
as $N \rightarrow \infty$.
We also look at this quantity in the interior of an infinite tree, and show
that a different transition takes place in this limit.

The plan of the paper is as follows: in Section \ref{sec:model} 
we introduce the 
model. In Sections \ref{sec:2satmom}, \ref{sec:3satmom} and \ref{sec:ksatmom} 
we calculate the moments of the 
number of assignments that are solutions averaged over all realizations for a randomly fixed boundary, 
for 2-SAT, 3-SAT and arbitrary $K$ 
respectively. In Section \ref{sec:rec_prob}, 
we calculate the probability that a 
given realization has at least one solution (or equivalently the fraction of realizations that have solutions), as a function of the depth of the tree. In Section \ref{sec:fp}, 
we carry out the fixed-point analysis relevant for the interior of an infinite tree.
We end with a summary and discussion in Section \ref{sec:summary}.

\section{The Model}
\label{sec:model}
%\subsection{Description}
\begin{figure}[tbp]
\centering 
\includegraphics[scale=0.6]{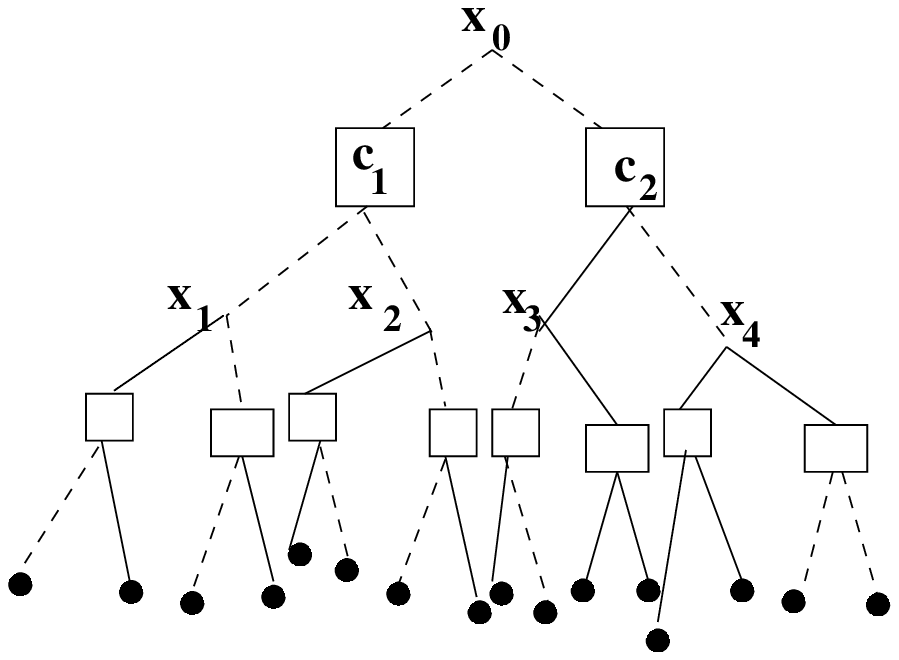}
  \caption{3-SAT on a rooted tree of depth $2$ and $d=2$. 
Only the clauses neighboring the root are labelled. 
The surface variables are depicted by a $\bullet$. Dashed/full lines between a variable and a clause indicate that it is negated/un-negated.}
  \label{fig:sat}
\end{figure}
We define the K-SAT problem on a tree as follows. Consider a
regular $d$-ary tree $T$  in which every vertex has exactly $d$
descendents.  The root of the tree $x_0$  has degree $d$ and its $d$ edges
are connected to function nodes $\{ c_1,c_2...,c_d\}$. Each function
node has degree $K$, and each of its $K-1$ descendents
$\{x_i=x_1,x_2,....x_{k-1} \}$ is the  root of an independent tree
(see Fig. \ref{fig:sat} ). Hence the root has a degree $d$ while all
the other vertices on the tree (except the leaves which have a degree $=1$) have a  degree $d+1$.  Each function
node is associated independently with a clause
$\phi(x_0,x_1,...,x_{k-1})$,  where the vertices $x_0,x_1..$ are the
neighboring vertices of the function node, joined to the function node
by a dashed or solid edge indicating whether the corresponding vertex
is negated or not in the clause. We consider only the case where the
vertices can take one of two values $0$ or $1$ and the case when every
function node is given by $\phi = \ell_0 \bigvee \ell_1
\bigvee....\ell_{k-1}$. Here $\ell_i$ is one of the two literals $x_i$ or $\overline{x_i}$,   
depending on whether $x_i$ is joined to the function node by a
dashed or a solid line (Fig. ~\ref{fig:sat}).

An assignment $\sigma$ of all the variables on the tree 
(barring the surface variables which take fixed values, see below) 
is a solution iff $\phi =1$
for all the  clauses on the tree. One configuration of dashed and
solid lines on the tree defines a realization $R$.

For a random K-SAT problem, {\i.e.} a K-SAT problem on a random graph,
the variable $\alpha = M/N$  where $M$ is the total number of clauses
and $N$ is the total number of variables (vertices), is a meaningful
quantity. As a function of this one can ask, for example, how 
the moments of the {\it total} number of  solutions (averaged over 
all realizations) scale with  $N$.

To ask the same question meaningfully here on a  tree, it is usual  to
fix the values of the variables on the surface of the tree. If we consider the
surface variables to have depth $0$, then we can denote this condition by
$\sigma(0)=L$, where $\sigma(0)$ is the assignment of the variables at
the $0^{th}$ depth and $L$ signifies a particular assignment for the
variables at this depth. Variables removed from the surface by one 
function node (or one level) are at depth $1$ and so on. 
The tree is said to have a depth 
$n$ if the root is $n$ levels away from the surface.

If the surface variables are fixed, then its easy to check that $M(n)=
d N(n)$, where $M(n)$ and $N(n)$ are respectively,  the total number
of clauses for a tree of depth $n$ and the total number of vertices 
($d=1$ if the surface variables are left free).
So $\alpha$ is the equivalent of $d$ on a tree with fixed boundary 
conditions.

Let us denote the total number of solutions (a sum over all $\sigma$
which are solutions) for a particular realization of a tree of depth
$n$ and  a specific boundary condition $L$ as $Z_R(L,n)$. This is a
stochastic variable which varies  from realization to realization as
well as from one boundary condition to another. The first moment of
this quantity, averaged over all realizations (for a fixed 
boundary condition), is trivially computed and is equal to 
$((\frac{2^{K}-1}{2^{K}})^{d}2)^{N(n)}$ (we derive this later in Section \ref{sec:2satmom}).  
In this case, if $(\frac{2^{K}-1}{2^{K}})^{d} 2 > 1$ , then the number of solutions
grows as the tree gets bigger and if $(\frac{2^{K}-1}{2^{K}})^{d} 2 <
1$ , then the number of solutions decreases as the number of variables
grows. The {\it critical} point of the first moment is therefore at
$(\frac{2^{K}-1}{2^{K}})^{d_c} 2 =1$ . This  gives $d_c =
\frac{-\log(2)}{\log(1-2^{-K})}$. Note that, from simple considerations,
it is easy to see that even the random K-SAT has the same
expression for the first moment as
on the tree, with $d$ replaced by $\alpha$. This expression for the
first moment gives an  annealed approximation for $\alpha_c(K)$.

We are henceforth interested in also estimating the higher moments for
$Z_R(L,n)$.  Before writing down the recursions for this quantity on the
tree however, we introduce a little more notation.  By $C^{s}(x_i)$
we denote all the clauses which are neighbours of  variable $x_i$ and
are satisfied by  it. Similarly, by $C^{u} (x_i)$ we denote all the
clauses which are neighbours of variable $x_i$ but  are not satisfied
by it. Let $F_R(L,n)$ and $G_R(L,n)$ denote the number of solutions
for a tree of depth $n$ (for a given realization and boundary
condition $L$) in which the root takes the value $0$ and $1$
respectively.

\section{Recursion Relations on the Tree for 2-SAT}
\label{sec:2satmom}
\begin{figure}
    \centering \includegraphics{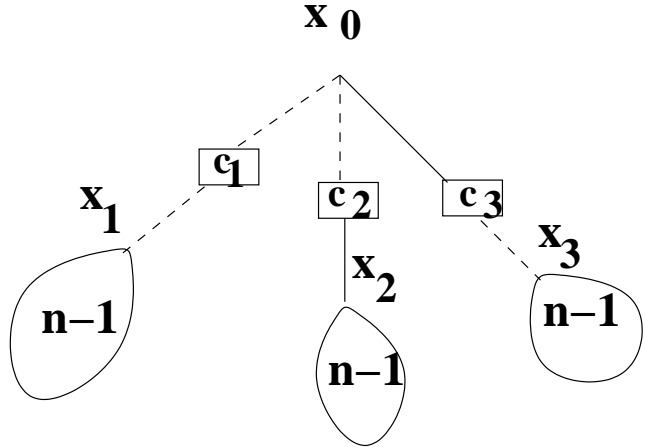}
    \caption{Schematic diagram of the tree for 2-SAT. Variable $x_0$- the root of a tree of depth $n$-  is connected through clauses $c_1$, $c_2$ and $c_3$ to variables $x_1$, $x_2$ 
and $x_3$ respectively, each of which is in turn the root of a tree of depth $n-1$. }
    \label{fig:treeschema}
\end{figure}

We can write the exact recursion equations on the tree for the stochastic variable $Z_R(L,n)$. We first look at the form of these recursions for 2-SAT. For ease of notation we henceforth omit the $L$ in the argument.  A {\it rooted} tree
of depth $n$ is generated by taking a root of degree $d$, then
picking, for all edges connecting the root to the function node,
independently the type of edge (dashed or solid)  and finally
attaching trees of depth $n-1$ to the other end of the function node
(again  via edges that are independently chosen to be dashed or
solid), see Fig. ~\ref{fig:treeschema}.

Let  $d_1$ be the cardinality of the set $C^{u}(x_0)$, when vertex
$x_0$ takes value $0$. Then  the vertices along the other edge of these
clauses can only take on the specific value satisfying the clause, for any realization of the
edge connecting them to the clause.
For those clauses satisfied by $x_0$ ($d-d_1$ of them; $d-d_1$ is just the cardinality of the set $C^{s}(x_0)$), the vertex at
the other edge of the clause is free to take any  value. So the number
of solutions for the tree of depth $n$, with root node taking value
$x_0=0$, for this specific  realization (and boundary condition), is
the product of the number of solutions of $d$ sub-trees of depth
$n-1$.

The recursion relation for $F_R$ is hence:
\be 
F_R(n) =\prod_{i=1}^{d-d_1} Z_{R_{i}}(n-1) \prod_{i=d-d_1+1}^{d} \left(
F_{R_i}(n-1) \eta_i + G_{R_i}(n-1) (1-\eta_i) \right)
\label{eq:F2sat}
\ee
where $R_i$ denotes the realization of each of the sub trees rooted in the
descendents of $x_0$ and $\eta_i$ is a stochastic variable which is
equally likely to take the value $0$ or $1$, depending on whether the edge joining the 
variable at depth $n-1$ to its clause is dashed or solid.
Similarly,
\be
G_R(n)  = \prod_{i=d-d_1+1}^{d} Z_{R_i}(n-1) \prod_{i=1}^{d-d_1}
\left( F_{R_i}(n-1) \eta_i + G_{R_i}(n-1)(1-\eta_i) \right)
\label{eq:G2sat}
\ee
and, 
\be 
Z_{R}(n)  = G_R(n)  + F_R(n)  
\label{eq:Z2sat}
\ee

We can define, $\beta_R(n) \equiv \frac{F_R(n)}{Z_R(n)}$ as the fraction of
solutions in which the root $x_0$ takes the value $0$ for a given
realization $R$ and a fixed but arbitrary boundary condition $L$. Note then that  
if, for a given $R$,
we selectively sum over those boundary conditions which result in a
certain $\beta_R(n)$, this gives the recursion for the residual
distribution at the root derived earlier in the context of tree
reconstruction  in \cite{mezard-montanari, bhatnagar-maneva}.

Our interest here however is to calculate the moments of $Z_R(n)$ from the
above recursion relations {\it by averaging over all}  realizations, for an
arbitrary boundary condition $L$.  Hence by $\langle F_R(n) \rangle$, we will denote 
an average over all realizations $R$ for a tree of depth $n$. It is easy to see that such an average is achieved at the root, by averaging over all possible $d_1$'s for a given $d$, where the probability that there are $d_1$ clauses
unsatisfied by the root is given by the binomial distribution $
P (d_1) = \frac{1}{2^d} \left(\begin{array}{c} d \\ d_1 \end{array}\right)$ .
Also note that since 
different branches  of the tree are independent of each other, we have,
\be 
\langle F_{R_i}(n) F_{R_j}(n)\rangle = \langle F_{R_i}(n)\rangle \langle F_{R_j}(n)\rangle =
\langle F_R(n) \rangle^2  
\ee
where $i$ and $j$ are variables along different branches.
In addition,
\be 
\langle F_{R_i}(n) \rangle = \langle G_{R_i}(n) \rangle 
\label{eq:symm}
\ee 
by symmetry. Infact this is true for any moment of $F$ (or $G$). 
It is also easy to see that
\be 
\langle \eta_i \rangle = 0.5 \ee \be \langle \eta_i (1-\eta_i)
\rangle = 0 
\ee 
and 
\be 
\langle \eta_i \eta_j \rangle = \langle
\eta_i \rangle \langle \eta_j\rangle = 0.25 
\ee 

Using the above equations, we can now solve Eq. \ref{eq:F2sat} to obtain the different
moments.

\subsection{First moment or average number of solutions}

The average number of solutions as discussed earlier, can be obtained
by just looking  at the probability of satisfying individual
clauses. In this section we average Eq. \ref{eq:F2sat} and make use
of Eq. \ref{eq:symm} to obtain (after removing subscripts for ease of notation):
\be  
\langle Z(n) \rangle  = 2 \langle F(n)\rangle 
\ee 
and
\be  
\langle F(n) \rangle = \frac{1}{2^d} \left( \langle Z(n-1)
\rangle + \langle F(n-1) \rangle \right)^{d} 
\ee

\be 
\langle F(n) \rangle =(\frac{3}{4})^d (\langle Z(n-1) \rangle)^d 
\ee 

which implies
\be 
\langle Z(n) \rangle  = 2 (\frac{3}{4})^d (\langle Z(n-1) \rangle)^d
\label{eq:av2sat}
\ee 
This is a simple recursion relation which we can solve
easily for any  boundary condition. The recursion relation
at the boundary is
\be 
\langle Z(1) \rangle  = 2(\frac{3}{4})^d 
\label{eq:bc2sat}
\ee 
and,
\be \langle F(1) \rangle  =(\frac{3}{4})^d 
\ee 
Eq. ~\ref{eq:bc2sat} follows from Eq. \ref{eq:av2sat} by noting that $Z(0) =1$.
%This is because the surface variables have
%fixed values. If, for a specific $\sigma(0)=L$, one joins $d$ such surface variables together to make a tree of
%depth $1$, then this tree, for a given realization has either $1$ or
%$0$ solutions when the root node takes the value  $0$, and the same
%when the root node takes the value $1$.  Averaging $F(1)$ over 
%all realizations implies estimating the probability that there exists at least one
%solution. This probability is simply $(\frac{3}{4})^d $. 
%By the same logic infact 
%\be 
%\langle F^m(1) \rangle  = (\frac{3}{4})^d 
%\ee 
Using the boundary condition Eq. \ref{eq:bc2sat}, we can solve the recursion relation Eq. \ref{eq:av2sat} to get the same result 
for the average number of solutions as mentioned in Section \ref{sec:model}. 

\subsection{Second moment}
 To get the second moment, note 
\be 
\langle Z^{2}(n) \rangle = 2 \langle F^{2}(n)\rangle + 2 \langle F(n)G(n) \rangle 
\ee 
From the recursion Eq. \ref{eq:F2sat} for the stochastic variable
$F_R(n)$, we have,
\be 
F^{2}_R(n) = \prod_{i=1}^{d-d_1} Z^{2}_R(n-1) \prod_{i=d-d_1+1}^{d} \left(
F_R(n-1){\eta_i}+ G_R(n-1)(1-\eta_i)\right)^2 
\ee 
This gives (again getting rid of subscripts),
\beq 
\langle F^2(n) \rangle &=& \frac{1}{2^d} \left(\langle Z^2(n-1) \rangle + 
\langle F^2(n-1) \rangle \right)^{d} \\
 &=& \frac{1}{2^d} \left( \langle 2 F^2(n-1) + 2 F(n-1)G(n-1)
\rangle + \langle F^2(n-1) \rangle \right)^{d} \\  
&=& \left((3/2) \langle F^2(n-1) \rangle + \langle F(n-1)G(n-1) \rangle
\right)^{d} 
\eeq 
Similarly the equation for $ \langle F(n)G(n)\rangle$ is:
\be 
\langle F(n) G(n) \rangle = \left( \langle F^2(n-1)\rangle + \langle F(n-1)G(n-1)\rangle
\right)^{d} 
\ee 
These two coupled equations need to be solved
to get the second moment. The boundary conditions are,
\be 
\langle F^2(1) \rangle=(\frac{3}{4})^d 
\ee 
which we get by noting that $\langle F^2(0) \rangle = 1/2$ and $\langle F(0)G(0) \rangle = 0$. Similarly, 
\be 
\langle F(1)G(1) \rangle = (\frac{1}{2})^d. 
\ee 
%We get the latter by estimating the probability
%that we have a solution {\it both} when the the root takes value $1$
%as well as $0$ (in all other cases, the value of the random variable
%$F_R(1)G_R(1)$ is $=0$). The root is free to take either value {\it
%only}  when all the $d$ surface variables are satisfying their
%clauses. This happens with probability $1/2^d$ since for any value of the surface variables,
%the probability of a realisation (of solid or dashed lines) which results in all $d$ of them satisfying their respective clauses is 
%$1/2^d$.

To solve the
coupled recursion equations, let us define the ratio 
\be 
r_n \equiv \frac{\langle F_R(n) G_R(n) \rangle}{\langle F^2_R(n)\rangle} 
\ee
Then we have the following equation for $r_n$. 
\be 
r_n =\frac{\left[ 1+ r_{n-1} \right]^d}{\left[ 3/2+ r_{n-1} \right]^d}
\ee
with boundary condition $r_1= (\frac{2}{3})^d $. 

For large $n$, $r_n$ reaches a fixed point and in that limit we can solve 
for the fixed point of this equation to get $r^\star = 0.78 (d=1), 0.576 (d=2), 
0.4(d=3)$. If we now approximate the equation for 
\be 
\langle F^2(n) \rangle \sim  \langle F^2(n-1)
\rangle^d (3/2+ r^\star (d))^d 
\ee
We can solve this to get (for $d > 1$),
\be 
\langle F^2(n) \rangle \sim \left[\frac{3}{4} \left( 3/2 + r^\star(d) \right)^{\frac{1}{d-1}}
\right]^{d^n} 
\ee
If the term in the brackets is $< 1$, the second moment decreases with system size. 
If it is $>1$, then the second moment  increases with system size. 
The ``critical'' value lies between $d \sim 3.1$ and $3.2$. 
To get the value of this threshold more precisely, we need to  solve the equation  
\be
\langle F^2(n) \rangle \sim  \langle F^2(n-1) \rangle^d (3/2+
r_{n-1}(d))^d
\ee
 exactly. Solving it numerically we get the critical
value of $d$ to lie between $3.06$ and $3.07$.

We can follow this procedure for any moment though the number of coupled recursions that have to be solved simultaneously, increase with the order of the moment. 

%\subsection{Third moment}

%We have, 
%\be 
%\langle Z^{3}(n) \rangle = 2 \langle F^{3}(n)\rangle
%+ 6 \langle F^{2}(n)G(n) \rangle 
%\ee.

%It turns that the
%recursion relations for the two quantities in the RHS, have the same
%form.
%\be 
%\langle F^{3}(n)\rangle = (3/2)^d \left( \langle
%F^{3}(n-1)\rangle + 2 \langle F^{2}(n-1)G(n-1) \rangle \right)^d
%\ee
%and,
%\be 
%\langle F^{2}(n)G(n)\rangle = \left(\langle F^{3}(n-1)\rangle + 
%2 \langle F(n-1)^{2}G(n-1) \rangle \right)^d 
%\ee
%This implies that 
%\be 
%r_n= \frac{\langle F^{2}(n-1)G(n-1) \rangle}{\langle F^{3}(n)\rangle} = (2/3)^d
%\equiv  r^\star 
%\ee
%Putting this back in the equation for
%$\langle F^{3}(n)\rangle$, we get, 
%\be 
%\langle F^{3}(n)\rangle \sim  \left[ \frac{3}{4} \left( 3/2 + 3 (2/3)^d
%\right)^{\frac{1}{d-1}} \right]^{d^n}
%\ee
% This gives a critical value $d_c$ between  $\sim 3.6$ and $3.7$.

\section{Recursion Relations for $K=3$}
\label{sec:3satmom}
The recursion relations for higher $K$ though slightly more
complicated, follow the same logic as for $K=2$. We carry out the 
computation for $K=3$ here. Now the recursion
relation for $F_R(n)$ is:
\beq
 &F_R (n)& = \prod_{i=1}^{d-d_1} Z_{R_{i1}} Z_{R_{i2}} \nonumber \\
&\prod_{i=d-d_1+1}^{d}& \left[Z_{R_{i1}}  Z_{R_{i2}}   
 -  \left( F_{R_{i1}} \eta_{i1} + G_{R_{i1}} (1-\eta_{i1}) \right) \left( F_{R_{i2}} \eta_{i2} + G_{R_{i2}} (1-\eta_{i2}) \right) \right] 
 \label{eq:F3sat}
\eeq 
where we have removed the dependence on $n-1$ in the LHS for ease of presentation. We can also write a similar equation for $G$. The $\eta$'s are the same
as appear earlier. The $i1$ and $i2$'s  signify the two variables at the
end of the same clause,  (see Fig \ref{fig:sat}).  Since they belong to two
different branches, their averages still decouple. The second term
is the counterpart of the term appearing in 2-SAT along the branches
where the root was not satisfying the clause (or link in 2-SAT). In
the case of 3SAT, if the root does not satisfy one of it clauses, the
other two variables are {\it collectively} constrained to not 
take $1$ out of the $4$ total assignments they could have had. Which
specific assignment is forbidden depends on the realization.

\subsection{First moment or Average number of Solutions} 
Again removing subscripts, 
\be  
\langle F(n) \rangle = \frac{1}{2^d} \left( \langle Z(n-1)\rangle^{2}
+ \langle Z(n-1)\rangle^{2} - \langle F(n-1) \rangle^{2} \right)^{d}
\ee

\beq 
\langle F(n) \rangle &=& \frac{1}{2^d} \left(2\langle Z(n-1) 
\rangle^2 - \langle F(n-1) \rangle^2 \right)^{d} \\
&=& (\frac{7}{8})^d (\langle Z(n-1) \rangle)^{2d} 
\eeq 
implying
\be 
\langle Z(n) \rangle  = 2(\frac{7}{8})^d (\langle Z(n-1) \rangle)^{2d} 
\label{eq:av3sat}
\ee 
As before, we need to solve the recursion relations keeping the boundary conditions in
mind, The recursion relation at the boundary is 
\be 
\langle Z(1) \rangle  = 2 (\frac{7}{8})^d \ee and,
\be 
\langle F(1) \rangle= (\frac{7}{8})^d 
\ee 
We can now solve Eq, \ref{eq:av3sat} to get the result for the annealed average, already mentioned in section ~\ref{sec:model}.

\subsection{Second moment}

Squaring Eq. \ref{eq:F3sat} and taking averages, we get, 
\beq 
\langle F^{2}(n) \rangle &=& \langle \prod_{i=1}^{d-d_1} Z_{i1}^{2}(n-1) 
Z_{i2}^{2}(n-1) \\ &\prod_{i=d-d_1+1}^{d}& \left( Z_{i1} Z_{i2} - \left(
F (n-1){\eta_{i1}}+ G(n-1)(1-\eta_{i1})\right) \left( F
(n-1){\eta_{i2}}+ G(n-1)(1-\eta_{i2}) \right) \right)^2 \rangle
\nonumber
\eeq 

Simplifying, we get,

\beq 
\langle F^2(n) \rangle &=& \frac{1}{2^d} \left( \langle Z^2(n-1) \rangle^{2} + \langle Z^2(n-1) \rangle^{2}
 + \langle F^2(n-1) \rangle^{2} -2 \langle F(n-1) Z(n-1)\rangle^{2} \right)^{d} \\   
&=& \frac{1}{2^d} \left( 7 \langle F^2(n-1) \rangle^{2} + 6 \langle F(n-1)G(n-1) \rangle^{2} + 12
\langle F(n-1)G(n-1) \rangle \langle F^2(n-1) \rangle \right)^{d}
\eeq
Similarly the equation for $\langle F(n)G(n) \rangle$ is 
\be 
\langle F(n) G(n) \rangle  = \left( 3 \langle F^2(n-1) \rangle ^{2} +  
3 \langle F(n-1)G(n-1) \rangle ^{2} + 6 \langle F(n-1)G(n-1) \rangle \langle F^2(n-1) \rangle \right)^{d} 
\ee

Again defining the ratio of $\langle F(n)G(n) \rangle$  to $\langle
F^2(n) \rangle$ as $r_n$, we get 
\be
\label{eq:fp3SAT}
r_{n+1} = \left[\frac{3+3r_n^{2}+6r_n}{7/2+3r_n^{2} +6r_n}\right]^{d}
\ee 
with $r_1 = (6/7)^{d}$. 

Going through the same
procedure as before, we find that the second moment diverges for  $d< 7.16$. 
In comparison, the first moment diverges for $d<5.19$.

\section{Recursion Relations for arbitrary $K$} 
\label{sec:ksatmom}
Similarly, we can easily write the recursions for $F_R(n)$ for any $K$.
Skipping details, the equation for the $r_n$'s  and for $\langle
F^2(n) \rangle$ for arbitrary $K$ are: 
\beq 
r_{n+1} = \left[\frac{(2^{K-1}-1) + g(r_n)}{(2^{K-1}-0.5) + g(r_n)} \right]^{d}
 \eeq 
where 
\be 
g(r_n)=  (2^{K-1}-1) \sum_{i=0}^{K-2} \left(\begin{array}{c} K-1 \\ i \end{array}\right) r_n^{K-1-i} 
\ee
and 
\beq
\langle F^2(n) \rangle &=& \langle F^2(n-1)
\rangle^{(K-1)d} \left( (2^{K-1}-1/2) + g(r_{n-1}) \right)^{d} 
\eeq 
These expressions can be simplified a bit. We can write
\beq 
r_{n+1} = \left[\frac{(1+r_n)^{K-1}}{\beta_K -1 +
(1+r_n)^{K-1}} \right]^{d} 
\eeq 
defining $(2^{K-1}-0.5)/(2^{K-1}-1) \equiv \beta_K$
 
 and 
\beq 
\langle F^2(n) \rangle &=& \langle F^2(n-1) \rangle^{(K-1)d}
(2^{K-1}-1)^{(K-1)d}\left(\beta_K-1 + (1+r_{n-1})^{K-1} \right)^{d} 
\eeq 
Solving these numerically for different $K$, we get the
numbers in the third column of table \ref{tab:c}.
\begin{table}
\caption{The critical values of $d$ for the first and second moments on the rooted tree, determined
by numerically solving the recursion relations derived in the text. The fourth column contains an estimate of the critical value from the expression for the second moment in Eq. \ref{eq:overlap}.}
\label{tab:c}
	\centering
		\begin{tabular}{|l|l|l|l|} \hline
		$K$ &  $\langle F \rangle$ & $\langle F^2 \rangle$ & $\langle Z_o^2 \rangle$   \\
		\hline  $2$ & $2.41$ & $3.06$ & $2.9$ \\ \hline  $3$ &
		$5.19$ & $7.16$ & $5.89$ \\ \hline $4$ & $10.74$ & $15.24$ & $11.6$  \\
		\hline $5$ & $21.832$ & $31.34$ & $22.8$   \\ \hline $6$ &
		$44.014$ & $63.52$ & $45.$  \\ \hline $7$ & $88.376$ &
		$127.86$ & $89.5$   \\ \hline $8$ & $177.099$ & $256.51$ & $177.6$   \\
		\hline
		\end{tabular}
\end{table}

That different moments pick out different critical values of $d$, is not confined
to the tree. We argue below that the K-SAT on a random graph, behaves similarly.

As mentioned earlier the behaviour of the first moment on the random graph is identical to that
on a tree from simple considerations.
The second moment, while not calculable in closed form for 
a random graph can be written in terms of the fraction of realizations that  
a pair of assignments are both simultaneously solutions of.
The probability of two assignments simultaneously being a
solution for a given realization depends on the overlap between the
two assignments. If two assignments have an overlap $pN$ ($0\leq p
\leq 1$), the probability that both are solutions is ${f(p)}^M 
\equiv (1-2^{1-k}+2^{-k}p^{k})^M $ where $M$ is the  total number of
clauses. The exact expression for the second moment is thus simply,
the number of pairs of assignments with overlap $pN$ times the
probability that such a pair are both simultaneously solutions for a
realization \cite{troyansky,achlioptas1}.
\be
 \langle Z_o^2 \rangle = 2^N \Sigma_{z=0}^{N} \left(\begin{array}{c}
N \\ z \end{array}\right)  f(p=z/N)
\ee

Following Achlioptas {\it et al }\cite{achlioptas1} and using
the leading-order  approximation $\left(\begin{array}{c} N \\ z \end{array}\right) =
\left(p^p (1-p)^{1-p}\right)^{-N} poly(N)$, the term which
contributes the maximum to the above sum is 

\be
\label{eq:overlap}
2^N \left( max_p \left[ \frac{{f(p)}^{\alpha}}{\left(p^p
(1-p)^{1-p}\right)} \right]\right)^N  poly(N)
\ee .

\begin{figure}
    \centering 
     \includegraphics[scale=0.6]{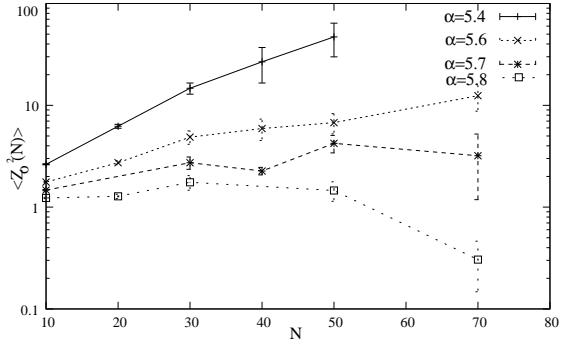}
    \caption{Second moment for random $3$-SAT as a function of
$N$ for different values of $\alpha$ }
    \label{fig:second}
\end{figure}

For our purposes its easy to see, by plotting the term which is 
exponentiated in $N$, as a  function of $p$, that there is a value of $\alpha$
above which the maximum value (which is an estimate of the second moment) 
is less than $1$. The numbers at
which this happens for different $K$ are reported in the fourth column of table
~\ref{tab:c}. We have also 
estimated the critical value using simulations.
For example, Fig.
\ref{fig:second} plots the second moment as a function of $N$ for various  
values of $\alpha$ for random $3$-SAT. We find that the
value of $\alpha $ for which the second moment starts  decaying with increasing $N$ is 
$~5.7 \pm 0.1$. Hence, at least for random 3-SAT, 
the value of the critical point for the second moment obtained from the overlap 
function is close to the numerical estimate.
Simulations for higher moments near the 
critical point are harder to do since the fraction  of
successful realizations decays exponentially with $N$. 
For example, at $\alpha=5.7$ and $N=70$, only $208$ 
out of the $10^6$ randomly chosen
realizations had  solutions. However we expect the higher 
moments to behave similarly, see for {\it e.g.} \cite{troyansky}
where  expressions for the third and fourth moments 
are also evaluated for random 3-SAT.

%{\textcolor{red}{As we see the numbers are different from those in Table
%\ref{tab:c}. There is also a qualitative difference, in that the
%difference between the 'critical' values for the moments increases
%with $K$ upto $K=8$ beyond which it decreases. Whereas in Table
%\ref{tab:c}, the difference only increases with $K$.

The quantitative differences in the values of the critical point for the second moment in between 
our model  and random $K$-SAT can be explained in terms of the boundary conditions of the tree
graph. For the tree graph, the probability that two assignments are
simultaneously solutions needs to be re-written for the nodes near
the boundary.  If we make the simple assumption that all nodes are
nodes at depth $1$, then we should replace $f(p)$ by $g(p)=
(1-2^{1-k}+2^{-k}p)$ in the expression for the second moment in Eq. \ref{eq:overlap}. 

The critical values of $d$ for the different moments, indicate the 
heterogeneity of different realizations.
Even at very large values of $d$ (or $\alpha$), 
there exist (an exponentially rare number of) 
realizations which by having an exponentially large 
number of solutions, contribute to the corresponding moment.
The critical values for different moments, 
also provide bounds on the solvability 
transition. For {\it e.g.}, the critical value for the first moment 
gives a  simple upper bound on the solvability transition for any $K$.
The ratio of the square of the first moment to the second moment 
provides a lower bound on the probability that solutions exist, and is the starting point for the weighted second moment method \cite{achlioptas_nature}. 
However to gain a better insight into the solvability transition, 
it is more useful to look at
how the fraction of solvable realizations changes as the number of constraints
increases. This quantity
can again be exactly calculated on a
rooted tree and we carry out this calculation in the next section.

\section{Exact recursion relations on the tree for the probability that 
a variable can take $2$, $1$ or $0$ values for arbitrary $K$}
\label{sec:rec_prob}

We would like to estimate the probability that a 
realization has no solution.
Such a realization is one for which not a single assignment of the
variables provides a solution. This  can happen if there is even a
single variable on the graph, which, whether it takes the value $0$ or
$1$, causes atleast one clause to be unsatisfied. Such a variable then
is a variable that can take $0$ values by our definition, and a
realization that is not solvable has at least one variable of this
type. 
 On the tree graph, we can define the probabilities of a variable
taking $0$, $1$ or $2$ values on the corresponding subtree.   
We define $p_i(0)$ as the conditional probability for a variable 
$x_i$ to cause a contradiction, in the subtree of which it is the root, given that all 
the other variables in the subtree can take at least $1$ value. We can then estimate the probability of a realization {\it having} 
a solution (or the fraction of realizations that have solutions) by  calculating
the quantity $\Pi_{x_i} (1-p_i(0))$ where the product is over all the
variables in the graph. The tree structure of the graph, gives us a
way of calculating $p_i(0)$ through recursions.

We define below
some of the quantities in terms of which these recursions are written.
We define $P_n(0)$ to be the probability that 
(or the fraction of realizations in which) a variable at depth $n$
can neither take the value $0$ nor take the value
$1$, without causing a contradiction in its subtree.
Note that because of the tree structure, all variables $x_i$ at depth $n$
will have the same probability $P_n$. The probability that a
variable at depth $n$, can take only one of the two values $0$ or $1$ is
defined to be $P_n(1)$. Similarly  the probability that a variable at
depth $n$ can take both values is $P_n(2)=1-P_n(0)-P_n(1)$.   As before, in all that follows,
we consider the boundary variables to have
depth $0$. In the sections below, we describe how to
set up the recursions for 2-SAT and then for a general K-SAT. 

We would like to calculate $P_{n+1}(0)$ and $P_{n+1}(1)$ for variable
$x_0$ (assuming it is at depth $n+1$), given these quantities for its
descendents.  Let us consider $P_{n+1}(0)$ first for the 2-SAT
problem. Assume variable $0$ has a degree $d$ (by definition) and
assume it is not negated on $d_1$ of these links. Variable $x_0$ will
not be able to take the value $0$ in the case when {\it at least} one
of the $d_1$ links is {\it not} satisfied by the variable at the other
end. In this case there will be at least one unsatisfied clause if
variable $x_0$ takes the value $0$. Similarly, at least one of the $d-d_1$
links along which variable $x_0$ {\it is} satisfying its link, should also
not be satisfied by the variable at the other end. This latter condition
implies that variable $x_0$  cannot take  the value $1$ either. It
is easy to see that averaging over all realizations at depth $n+1$
implies averaging over all values of $d_1$ as before, and averaging
over all realizations  at depth $n$. It is important to note however
that all the realizations at depth $n+1$ are only built up from those
realizations at depth $n$ that do have solutions.

Putting all this together the recursion relations for $2$-SAT are:
\beq
\label{eq:2satprob}
P_{n+1}(0) &=& \sum_{d_1} \frac{1}{2^d} \left(\begin{array}{c} d \\ d_1 \end{array}\right) \left[ 1-\left(1- \left( \frac{P_n(1)}{2 (1-P_n(0))}\right)\right)^{d_1}\right] \left[ 1-\left(1- \left( \frac{P_n(1)}{2 (1-P_n(0))}\right)\right)^{d-d_1}\right] \nonumber \\ 
&=& 1-2 \left( 1-0.5 \left( \frac{P_n(1)}{2 (1-P_n(0))}
\right)\right)^d + \left(1-\left( \frac{P_n(1)}{2 (1-P_n(0))}
\right)\right)^d \\  
P_{n+1}(1) &=& 2 \left( 1 - 0.5 \left(
\frac{P_n(1)}{2 (1-P_n(0))} \right) \right)^d - 2 \left(1- \left(
\frac {P_n(1)} {2 (1-P_n(0))} \right)\right)^d 
\nonumber
\eeq
The term $P_n(1)/(2 (1-P_n(0))) \equiv Q_n(1)$ which appears in the
above equations is just the conditional probability that a depth
$n$ variable cannot satisfy its link above (to depth $n+1$), given
that it has to be able to take at least one value (which satisfies the
sub tree of which it is the root). The only way the latter can happen
is if the one value it can take, does not satisfy the link above (this
happens with probability $1/2$ in our case). If the variable can take two
values,  we can always find one value which satisfies the link above
for any realization. We can iterate these equations beginning
with the boundary conditions 

\beq
\label{eq:bc}
P_1(1)&=& 2(0.75)^d- 2(0.5)^d \\
 P_1(0) &=& 1+(0.5)^d - 2(0.75)^d
\nonumber 
\eeq 

The boundary conditions are easily obtained by
setting $P_0(1)=1$ in Eqns \ref{eq:2satprob}. 
We can hence obtain $P_n(0)$ for any depth
$n$. The probability that a realization has a solution is then (as
mentioned earlier) $\Pi_n (1-P_n(0))^{g(n)}$ where $g(n)$ is simply
the number of variables at depth $n$. If $P_n(0) \neq 0$ 
then the above product decays exponentially to $0$ 
with increasing tree size.
We see from Eq. ~\ref{eq:2satprob} that 
it is when $P_n(1)$ takes a non-zero value that the fraction of 
realizations that have no solutions also becomes non-zero. 

For a tree graph, since the boundary vertices form most of the
graph, we can just take a look at equation \ref{eq:bc}. If we plot
$P_1(0)$ as a function of $d$, we see immediately that this is nonzero
for any $d>1$ and hence, the fraction of realizations that do not
have solutions is non-zero above $d=1$. This conclusion is true, even
if we calculate $\Pi_n (1-P_n(0))^{g(n)}$ using the recursions 
and calculating $g(n)$-the number of vertices at depth
$n$, for each depth. Hence for $d>1$ the probablity of having 
a realization with solutions goes  down exponentially with $N$. For example, 
for $d=3$, $K=2$ and $n=4$, we find from exact enumerations, that out  
of $10^7$ randomly generated instances, only $77$ have solutions. 
Note however that these $77$ suffice to make  the second
and higher moments still an increasing function of $N$, as we saw in the 
previous section.

The recursion relations for 2-SAT are easily generalised to arbitrary
K.
\beq 
\label{eq:Ksatprob} 
P_{n+1}(0) &=& 1-2 \left(
1-0.5(Q_n(1))^{K-1}\right)^d + \left(1-(Q_n(1))^{K-1}\right)^d \\
P_{n+1}(1) &=& 2\left( 1 - 0.5(Q_n(1))^{K-1} \right)^d -
2\left(1-(Q_n(1))^{K-1}\right)^d 
\nonumber 
\eeq  
with the boundary conditions obtained by putting $P_0(0)=0$ and $P_0(1)=1$ as
before.  

We find that for all $K$, for $d>1$, the fraction of
realizations that have  solutions decays exponentially with $N$. 
Since this is also usually how the SAT-UNSAT transition is 
defined numerically, we can conclude that $\Pi_n (1-P_n(0))^{g(n)}$ 
is the order parameter for the solvability transition in our model.
The transition occurs at $d=1$ for all K when 
the boundaries are fixed randomly.
In comparison, for random 3-SAT the value of 
$\alpha$ (that we find from simulations) at which the fraction of successful
realizations starts decaying exponentially is  $\alpha=4.25 \pm 0.05$.
Hence both the behaviour of the moments as well as the 
the SAT-UNSAT transition, is qualitatively similar to random K-SAT, though
quantitatively, on the tree graph, these are influenced mostly by the
surface variables, as expected. Hence in the next section, we 
look at the behaviour of Eqns. \ref{eq:2satprob} and \ref{eq:Ksatprob} 
in the interior of the tree, for an infinite rooted tree.

\section{Fixed point analysis}
\label{sec:fp}
%In our model, if we do not fix the boundary, the ratio of number of clauses to number 
%of variables ($\alpha$) approaches 1 independent of the degree $d$ of the tree. As 
%shown in the previous section, boundary effects are very important. 
In this section we will try to get rid of boundary effects 
by looking deep within the tree. This is usually equivalent 
to doing a calculation on a Bethe lattice, in which case,
all vertices on the tree should be equivalent and have 
the same coordination number.  On the rooted tree we have discussed
so far, the root has a coordination number $1$ less than the other sites.
However, from the structure of the recursions Eq. \ref{eq:Ksatprob},
for any vertex/variable on the tree, it is only $d$ of the connections 
(to descendents)
which determines the probability $P_n(0)$. Hence $P_n(0)$ at a depth $n$ 
is not changed when more levels are added to the tree. 
The same is not true for $P_n(1)$
which, at least when $P_n(0) \neq 0 $, needs to be corrected, for interior sites.
However this correction does not affect the transition that we discuss below.
We can hence hope that the fixed point of the recursions Eqns. ~\ref{eq:2satprob} 
and ~\ref{eq:Ksatprob} will give us an insight on how these probabilities behave in the 
interior of the tree, independent of boundary conditions.

%Hence we connect $(d+1) (K-1)$ rooted trees with $n$ depths 
%via $d+1$ clauses at a central site. Assuming that the central site is at depth $n+1$, 
%the probability that the central site can take no value ($P_{n+1}(0)$) or can take one 
%value ($P_{n+1}(1)$) will follow the following recursions relations respectively:}

%\beq
% P_{n+1}(0) &=& 1-2 \left(1-0.5(Q_n(1))^{K-1}\right)^{d+1} + \left(1-(Q_n(1))^{K-1}\right)^{d+1} \\
%P_{n+1}(1) &=& 2\left( 1 - 0.5(Q_n(1))^{K-1} \right)^{d+1} -2\left(1-(Q_n(1))^{K-1}\right)^{d+1} \nonumber 
%\eeq
%{\it where $Q_n(1)$ is the same quantity as defined in the previous section on rooted trees. 
%In the limit of $n \rightarrow \infty$, value of $1-p_{n+1}(0)$ will be the order parameter. In this limit 
%$Q_n(1)$ will approach a fixed point value, which can be substituted in equations above to get the 
%value of $1-p(0)$. Hence, we need to look at fixed point of equations for $Q_n(1)$ in order to 
%get rid of boundary effects.} 

Consider the
recursions (Eq. \ref{eq:2satprob}) for $2$-SAT first.  Noticing that it is the
quantity $Q_n(1)$ which appears on the RHS of these equations, we can
rewrite the recursions as 
\be
Q_{n+1}(1) = \frac{\left[1-0.5 Q_n(1)
\right]^d - \left[ 1- Q_n(1)\right]^d}{{2 \left[1-0.5 Q_n(1) \right]^d
- \left[ 1- Q_n(1)\right]^d}} 
\ee

If the above map has a fixed point, then the value of $Q_n(1)=
Q_{n+1}(1)= Q^*$ and $Q^*= f(Q^*)$ where $f(Q^*)$ is the function on
the RHS and $0<Q^*<1$.

\begin{figure}
    \centering \includegraphics[scale=0.5,angle=270]{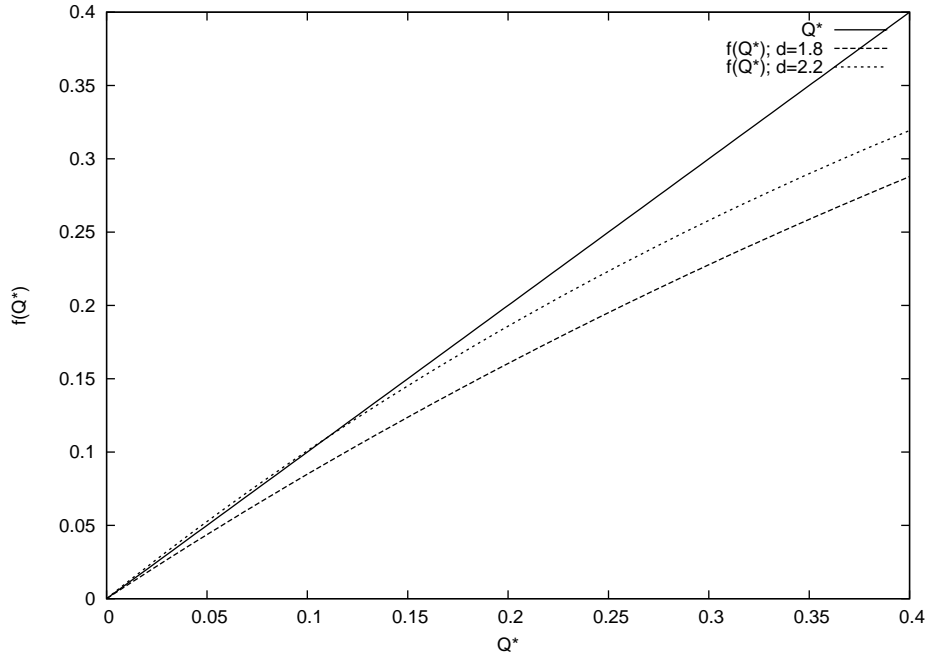}
    \caption{Fixed points for $d<2$ and $d>2$ for 2-SAT}
    \label{fig:rec1}
\end{figure}

We can look for this graphically as shown in Fig ~\ref{fig:rec1}. For
$d<2$, there is only one solution to the  equation $Q^*= f(Q^*)$ and
this lies at $Q^* =0$. For $d>2$ there are two solutions, one at $Q^*
=0$ and the other at some non-zero $Q^*$. 
If $Q^*=0$, clearly $P_n(0) = 0$, for large
$n$. This implies that (assuming all $N$ nodes in our system are deep
in the tree), the probability that a realization has a solution  ($\Pi
(1-P(0))$) is $=1$. Equivalently if $Q^* \neq 0$, this probability 
vanishes as $N \rightarrow \infty$.
Note that the non-zero value of $Q^*$ develops
continously from $Q^*=0$. In other words, for $d$ only very slightly
larger than $2$,  the non-trivial solution is only very slightly
larger than $0$.  In addition, from the shape of the function $
f(Q^*)$ we see that no matter what the boundary conditions, 
the non-trivial fixed point is  {\it always} reached for
$d>2$. Hence for 2-SAT, in the interior of an infinite tree, 
a real transition (the solvability transition)
occurs at $d=2$. 

On general grounds \cite{baxter}, 
we expect that the above results are comparable with
those on a regular random graph with coordination number $\alpha=(d+1)/2$ and 
hence a transition at $d=2$ in the interior of a tree should correspond to
an $\alpha_c=1.5$. It is interesting to note however that if
we use instead the relation $\alpha=d/2$ (as used in the 
correspondence between the cavity method and tree-reconstruction)
this gives the solvability
transition for random 2-SAT to lie at $\alpha=1$ which is an exact and known
result.

%{\it The ratio of clauses to variables inside the uniform tree(on Bethe lattice) is
%$(d+1)/2$(cite baxter gujrati). This would imply that in the limit in which we ignore 
%the boundary conditions the value of $\alpha$ at which the transition happens on our 
%regular tree would be $3/2$. We can also try to make a connection between an infinite
%rooted tree and a random graph in order to get analogous value for the random $2$-SAT. 
%As discussed by Mertens, Mezard Zecchina (and also Mezard Montanari) the relation between 
%the $d$ and $\alpha$ for random $2$-SAT is $\alpha=d/2$. 
%This gives the solvability
%transition for random 2SAT to lie at $\alpha=1$ which is an exact and known
%result.}

Consider the case $K>2$ now.  As before, we can write \beq Q_{n+1}(1)
= \frac{\left[1-0.5 Q_n(1)^{K-1} \right]^d - \left[ 1-
Q_n(1)^{K-1}\right]^d}{{2 \left[1-0.5 Q_n(1)^{K-1} \right]^d - \left[
1- Q_n(1)^{K-1} \right]^d}} \eeq

\begin{figure}
    \centering \includegraphics[scale=0.5,angle=270]{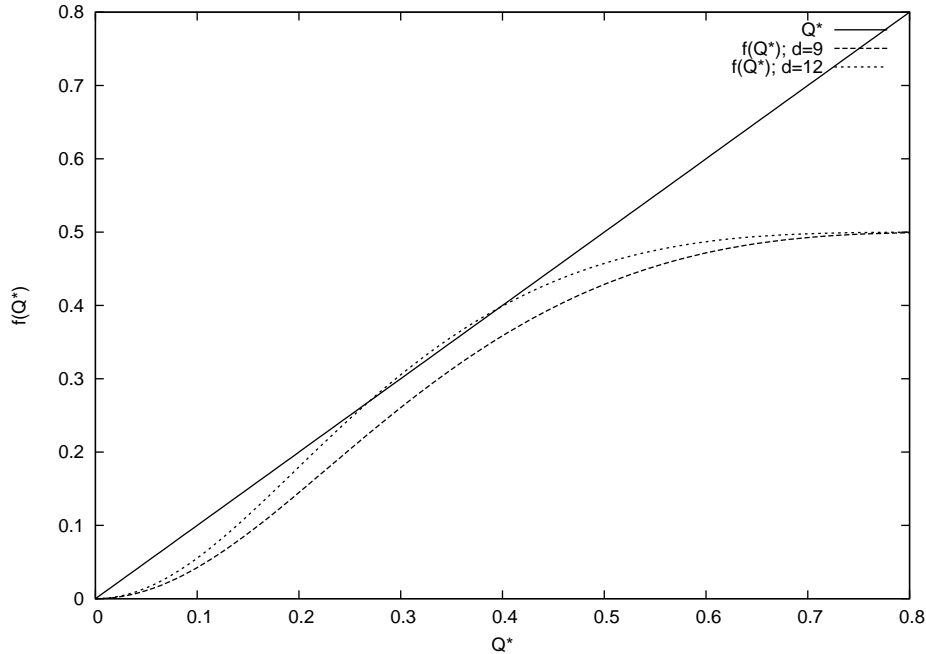}
    \caption{Fixed points for $d<11.5$ and $d>11.5$ for 3SAT}
    \label{fig:rec2}
\end{figure}

We can investigate the fixed points graphically as shown for $K=3$ in
Fig. ~\ref{fig:rec2}.  For $d<11.5$,  there is
only one solution to the  equation $Q^*= f(Q^*)$ and this lies at $Q^*
=0$ as before. For $d>11.5$ there are {\it three} solutions, one at
$Q^* =0$  and the other two at some non-zero $Q^*$.

Some important points of difference with 2-SAT are, that the non zero
values of $Q^*$ develop discontinously from $Q^*=0$. Also, from the
shape of the function $f(Q^*)$, the first ($Q^*=0$) and third
(non-trivial) value of $Q^*$ are  stable while the middle value is
unstable.  This is true for all $K \geq 3$. Hence from the shape 
of the function $ f(Q^*)$ we also see that boundary conditions play an 
important role for $K>2$.  The non-trivial solution is reached only if 
the boundary conditions are such as to make the value $Q_0(1)$ larger than 
the value of the central fixed point. 
This points to a first order transition for $K \geq 3$ 
as opposed to a continuous transition for $K=2$. This difference in the 
nature of the transitions at $K=2$ and $K>2$  is entirely
analogous to the problem on a random graph \cite{monasson}. 

For $K>2$ the correspondence between the average degree of a variable
on a regular random graph and that of a variable on the tree comes out 
to be $ \alpha= (d+1)/K$. 
%\footnote{This can be derived for general $K$-SAT tree by looking 
%at the expression for bulk free energy. This result is a generalisation of the well known 
%result for Cayley tree known as the homegenity condition (\cite{baxter,gujrati}).} 

In table ~\ref{tab:e} we report the value $d_c$ 
as obtained from the fixed point equations for our model, as well as both
$(d_c+1)/K$ and $d_c/K$. It is interesting to note that
these latter values are very close to the value of $\alpha_d$ obtained in the 
literature earlier ~\cite{mmz,montanari-semerjian}.

Note that in our case the value $d_c$ appears as that value of $d$ in the 
interior of an infinite tree, from which point onwards a 
solvability transition could take place, depending on the boundary values. 
It is only for 2-SAT that the transition actually does co-incide with 
$d$ taking the value $d_c$.

\begin{table}
\caption{The value of $d_c$ compared to $\alpha_d$ values obtained from \cite{mmz}.}
\label{tab:e}
	\centering
		\begin{tabular}{|l|l|l|l|l|} \hline
		$K$ & $d_c$ & $(d_c+1)/K$ & $d_c/K$ & $\alpha_d$ (from \cite{mmz})  \\ \hline  $2$ &
		$2$ & $1.5$ & $1$ & $1$ \\ \hline $3$ & $11.5$ & 4.166 & $3.83$ & $3.927$ \\ \hline
		$4$ & $32.6$ & $8.4$ & $8.15$ & $8.297$ \\ \hline $5$ & $80$ & $16.2 $ & $16.00$ &
		$16.12$ \\ \hline $6$ & $ 182$ & $30.5 $ & $30.33$ & $30.5$ \\ \hline $7$
		& $400 $ & $57.28 $ & $57.14$ & $57.22$ \\ \hline $8$ & $856$ & $107.13 $ & $107.00$ &
		$107.24$ \\ \hline
		\end{tabular}
\end{table}

\section{Summary and Discussion}
\label{sec:summary}
In this paper, we have studied the K-SAT problem on a rooted tree
and have solved it exactly for several quantities. A lot of progress 
in understanding the K-SAT problem has  been made using the cavity method \cite{mezard-science,mmz,braunstein} 
and a powerful heuristic, survey propagation(SP) 
\cite{sp,braunstein} has been developed using these concepts. The 
SP equations, which are the basis for the algorithm, 
are a set of coupled equations for the cavity bias surveys --messages 
sent from clauses to variables-- in terms of 
the probabilities of warnings  received by  the variable. 
The probability space over which these are computed is the 
space built by all SAT assignments each given equal probability 
in a typical satisfiable instance. It is conjectured that this 
solution space, for $\alpha > \alpha_d$, separates into many distant 
clusters, and hence the SP message along an edge gives the probability
that a warning is sent in a randomly chosen cluster. Under this assumption, the SP equations 
lead to coupled integral equations with a non-trivial fixed point, known as the 
'one step RSB' solution.  The same fixed point equations may also be obtained by  different means, by considering a reconstruction problem on a tree \cite{mezard-montanari}. The difference between a RS (replica-symmetric) solution 
and a 1RSB solution in the tree case, is due to the absence or presence of a 
re-weighting factor in the fixed point equation \cite{mezard-montanari}. This re-weighting factor may be thought of 
as the term in Eqns. \ref{eq:F2sat}
and \ref{eq:G2sat} (for e.g.), which involves the $\eta$ 's. Were
we to replace the $\eta$ 's simply by $1/2$ (their average value), 
then the recursion would give the number of solutions at the next level of the tree as simply the product 
over the previous level. The presence of the $\eta$ 's makes the recursion more non-trivial.
The recursions in Sections \ref{sec:rec_prob} and 
\ref{sec:fp} too, keep this re-weighting factor 

%We do not however need to impose the clustering condition because of the nature of the quantity we are looking at 
%(which is blind to clustering), namely the fraction of realizations in which a variable is constrained to take the same value 
%in {\it all} solutions (and hence all clusters).

Another difference in our work from the cavity method or SP is that
the marginal probabilities are
calculated level by level, instead of for each variable separately. 
This simplifies analysis, but makes our treatment valid strictly only for trees, unlike SP which is used to solve the SAT problem also on random graphs (though there are no guarantees of convergence in this case).

Our model shows many of the non trivial features of the full random K-SAT 
problem. To summarise, we find that different moments of the number of 
solutions start to decay at different values of $d$. These values of 
$d$ are much larger than the value at which the solvability transition 
occurs. Also the solvability transition in our model is continuous for 
$K=2$ and discontinuous for $K>2$, as in random K-SAT.
In addition the fixed point equations predict for $K>2$, a lower bound on the 
solvability transition in the interior of the tree.
These numbers if converted to equivalent $\alpha$ values on the 
random graph are very close to the values predicted for $\alpha_d$ 
in the literature for random graphs with an average node-degree $K\alpha$ (to make a more literal correspondence to our tree where every variable has the same degree, it would be interesting to compare with $\alpha_d$ values predicted for regular random graphs). We should note though that 
the existence of $\alpha_d$ is 
motivated in the literature 
by the structure of the space of solutions in random K-SAT,
while in this work, it arises as the point at which the recursions can have more than one solution, depending on boundary conditions. At this point, the fraction of realizations in which a variable is constrained to take the same value in all SAT assignments becomes non-zero.

We can redo the computation of Eqns. ~\ref{eq:Ksatprob} as well 
as the fixed point analysis, for other variations 
of the SAT problem such as regular random K-SAT, introduced in ~\cite{boufkhad}
and for which bounds on the threshold are derived in ~\cite{vish}.
Preliminary results from the fixed point analysis show that 
the fixed point equations have a similar behaviour as for K-SAT, but predict
smaller values of $d_c$ for the same $K$ ~\cite{inprep}, as indeed
is also the numerical prediction for the solvability transition 
for this problem \cite{boufkhad}.
It should also be possible to redo these calculations on a 
random tree (where the degree of each vertex is Poisson-distributed), 
though we do not expect the results to change qualitatively in this case.
Another interesting generalization of these calculations is to 
compute a more fine-grained quantity than $P_n(2)$, namely to compute
the probability that, the root takes one of the two values a certain fraction
$\beta$ of the times. It would be interesting to see if this quantity
undergoes a transition in which $\beta$ changes from essentially taking the value $1/2$ to having a non-trivial distribution. Note that a similar 
calculation for 
random tree ensembles, with however an average done over boundary conditions chosen uniformly 
over all satisfying assignments, is done in \cite{semer}.

\section{Acknowledgements}

We would like to thank Satya Majumdar for extremely useful discussions 
at the start of this project. We would also like to thank 
Elitza Maneva for a critical reading of the manuscript and Guilhem Semerjian and Martin Weigt for many insightful suggestions on the results as well as
on the improvement of the presentation. During most of this work, 
S.K was supported by the swedish research council.


\begin{thebibliography}{99}
%\setlength{\itemsep}{-1mm}
\bibitem{cook} S. Cook, Proceedings of the 3rd Anuual ACM Symposium on
Theory of Computing, Shaker Heights, Ohio, United  States, (ACM Press,
New York, 1971), p 151.
\bibitem{msl} D. Mitchell, B. Selman and H. Levesque, 
Proc. 10th Nat. Conf. Artif. Intel., 459 (1992).
\bibitem{kirkpatrick} S. Kirkpatrick and B. Selman, Science {\bf 264},
1297 (1994)
\bibitem{friedgut} E. Friedgut, J. Am. Math. Soc. {\bf 12}, 1017 (1999).
\bibitem{chvataletal} V. Chvatal and B. Reed, 33rd FOCS, 620 (1992); A. Goerdt, J. Comput. System. Sci., {\bf 53}, 469 (1996); W. Fernandez de la vega, 
Theoret. Comput. Sci. {\bf 265}, 131 (2001).
\bibitem{achlioptas_bounds} D. Achlioptas, 
Theoret. Compt. Sci. {\bf 265}, 159 (2001).
\bibitem{mezard-science} M. Mezard, G. Parisi and R. Zecchina, Science {\bf
297}, 812 (2002). 
\bibitem{mmz} S. Mertens, M. Mezard and R. Zecchina, Random
Structures and Algorithms, {\bf 28}, 340 (2006).
\bibitem{mezard-montanari} M. Mezard and A. Montanari, 
J. Stat. Phys. {\bf 124}, 1317 (2006).
\bibitem{derrida} B. Derrida, Phys. Rev. Lett. {\bf 45},
79 (1980);B. Derrida, Phys. Rev. B  {\bf 24}, 2613 (1981).
\bibitem{bhatnagar-maneva} N. Bhatnagar and E. Maneva, To appear in SIAM J. on Discreet Math., {\bf 25}, 854 (2011)
\bibitem{troyansky} L. Troyansky and N. Tishby, PhysComp96, Extended
abstract, May 1996.
\bibitem{achlioptas1} D. Achlioptas and C. Moore, SIAM J Comput. {\bf 36}, 740(2006).
\bibitem{achlioptas_nature} D. Achlioptas, A. Naor and Y. Peres, Nature {\bf 435},759(2005).

\bibitem{baxter} R. J Baxter, Exactly solvable models in statistical mechanics, 
Academic Press 
\bibitem{monasson} R. Monasson and R. Zecchina, Phys. Rev. Lett. {\bf
76}, 3881(1996); R. Monasson and R. Zecchina, Phys. Rev. E {\bf 56}, 1357(1997)
\bibitem{montanari-semerjian} A. Montanari, F. Ricci-Tersenghi and G. Semerjian, J. Stat Mech., P04004 (2008).
\bibitem{boufkhad} Y. Boufkhad, O. Dubois, Y. Interian and B. Selman, 
J. of Autom. Reasoning, {\bf 35}, 181 (2005).
\bibitem{vish} V. Rathi, E. Aurell, L. K. Rasmussen and M. Skoglund, SAT 2010, 264 (2010).
\bibitem{inprep} Draft.
\bibitem{semer} G. Semerjian, J. Stat. Phys.,{\bf 130}, 251 (2008)
\bibitem{braunstein} A. Braunstein, M. Mezard and R. Zecchina,  Random
Structures and Algorithms, {\bf 27}, 201 (2005).
 \bibitem{sp} M. Mezard and R. Zecchina,  Phys. Rev. E, 66:056126 (2002).
%\bibitem{mezard} M. Mezard, T. Mora and R. Zecchina,
%Phys. Rev. Lett. {\bf 94}, 197205(2005); D. Achlioptas and
%F. R. Tersenghi
%\bibitem{crawford} J. A. Crawford and L.D. Auton, Artificial Intelligence,
%{\bf 83} 31(1996)
%\bibitem{aurell} E. Aurell, U. Gordon and S. Kirkpatric
%\bibitem{cocco} S Cocco, R. Monasson, A. Montanari and G. Semerjian
%\bibitem{achlioptas2} D. Achlioptas and F. R. Tersenghi, STOC 06, Proceedings of the 
%thirty-eighth annual ACM symposium on Theory of computing 
%\bibitem{maneva} Elitza Maneva, Elchanan Mossel, Martin J. Wainwright, A new look 
%at survey propagation and its generalizations, Proceedings of the sixteenth annual 
%ACM-SIAM symposium on Discrete algorithms, January 23-25, 2005, Vancouver, British Columbia 
%\bibitem{montanari} A. Montanari, R. Restrepo and P. Tetali
%\bibitem{gujrati} P. D. Gujrati Phys. Rev. Lett. {\bf 53}, 2453(1984).


\end{thebibliography}
\end{document}